\documentclass[11pt,superscriptaddress,aps,prd,preprint]{revtex4}
\usepackage{amsmath}

\makeatletter
\usepackage[T1]{fontenc}
\usepackage{amsmath}
\usepackage{amssymb}
\usepackage{graphicx}

\usepackage{slashed}

\DeclareSymbolFont{extraitalic}      {U}{zavm}{m}{it}
\DeclareMathSymbol{\Qoppa}{\mathord}{extraitalic}{161}
\DeclareMathSymbol{\qoppa}{\mathord}{extraitalic}{162}
\DeclareMathSymbol{\Stigma}{\mathord}{extraitalic}{167}
\DeclareMathSymbol{\Sampi}{\mathord}{extraitalic}{165}
\DeclareMathSymbol{\sampi}{\mathord}{extraitalic}{166}
\DeclareMathSymbol{\stigma}{\mathord}{extraitalic}{168}

\newcommand{\bea}{\begin{eqnarray}}
\newcommand{\eea}{\end{eqnarray}}

\begin{document}

\title{Ricci dark energy in bumblebee gravity model}

\author{W. D. R. Jesus}\email[]{willian.xb@gmail.com}
\affiliation{Instituto de F\'{\i}sica, Universidade Federal de Mato Grosso,\\
78060-900, Cuiab\'{a}, Mato Grosso, Brazil}

\author{A. F. Santos}\email[]{alesandroferreira@fisica.ufmt.br}
\affiliation{Instituto de F\'{\i}sica, Universidade Federal de Mato Grosso,\\
78060-900, Cuiab\'{a}, Mato Grosso, Brazil}

\begin{abstract}

The Ricci dark energy is a model inspired by the holographic dark energy models with the dark energy density being proportional to Ricci scalar curvature. Here this model is studied in the bumblebee gravity theory. It is a gravitational theory that exhibit spontaneous Lorentz symmetry breaking. Then the modified Friedmann equation is solved for two cases.  In the first case the coupling constant $\xi$ is equal to zero. And in the second case a solution in the vacuum, where the bumblebee field becomes a constant that minimizes the potential, is considered. The coupling constant controls the interaction gravity-bumblebee. 

\end{abstract}

\maketitle

\section{Introduction}

General relativity describes gravitation at a classical level. Although it is a theory of gravity successful tested, it does not explain some observational data. Observational results lead to evidences for a phase of accelerated expansion of the current universe \cite{Riess, Permutter, Sp1, Sp2, Teg, Se,  Will2006a}. This acceleration may be explained by a new component called dark energy that corresponds to approximately 70\% of the energy content of the universe and its nature is still not clear at the present moment. The cosmological constant $\Lambda$ is the simplest candidate to describe dark energy. It is a fundamental ingredient of the $\Lambda$-CDM model \cite{Spergel:2003cb}, the most consistent model with the experimental observations. However, this model suffers with two major issues, the fine-tuning problem (the observed value of the cosmological constant is of the order of $10^{-120}$ smaller than the estimated vacuum energy in quantum field theory) and coincidence problem (in the current period of the universe history the values of the densities of dark energy and dark matter are of the same order of magnitude) \cite{Bull:2015stt}. 

An interesting approach to explain the late acceleration of the universe is the Ricci Dark Energy model (RDE) \cite{gaoRDE}. It is inspired on the holographic principle \cite{Susskind} and it considers that the dark energy density is proportional to the Ricci scalar curvature. The great advantage of this model is that it avoids both problems, the fine-tuning and coincidence problem, since dark energy density is not associated Planck scale but with cosmological scale. The holographic dark energy has been studied in various contexts. For example, the DGP braneworld with time varying holographic parameter has been investigated in \cite{GHAFFARI201976}, the plane symmetric modified holographic Ricci dark energy model in Saez-Ballester theory of gravitation has been analyzed in  \cite{RAO2018469}, the holographic dark energy in 5D Brans-Dicke theory has been considered in \citep{Salehi:2018czt}, the Ricci dark energy model with bulk viscosity has been studied in \cite{Singh2018}, the Ricci dark energy in Chern-Simons modified gravity has been presented in \cite{Silva2013}, among others.

Another problem evolving the general relativity is the lack of a quantum gravity theory. However, there are some attempts to construct a fundamental theory that unifies general relativity theory and the standard model of particle physics. This theory is expected to emerge at the Planck scale, $\sim 10^{19}\mathrm{GeV}$, where
small Lorentz violation effects may appear  \cite{PhysRevD.39.683, Kost1991}. Lorentz symmetry breaking  arises as a possibility in string theory \cite{PhysRevD.39.683}, noncommutative field theories \cite{Carroll} and  loop quantum gravity theory \cite{Gambini}. To study Lorentz violation consequences an extension of the standard model has been developed. The Standard Model Extension (SME)  \cite{SME1, SME2} is an effective field theory that contains the standard model, general relativity and all possible operators that break Lorentz symmetry. Here the bumblebee gravity is considered \cite{PhysRevD.39.683}. The bumblebee model is an effective theory of gravity in which spontaneous Lorentz violation is induced by a potential $V(B_\mu)$, where a vector field $B_\mu$ acquires a nonzero vacuum expectation value. Several applications with the bumblebee model have been done, such as,  traversable wormhole solution in the framework of the bumblebee gravity theory \cite{Ovgun:2018xys}, exact Schwarzschild-like solution \cite{Casana:2017jkc}, cosmological implications of Bumblebee vector models \cite{Capelo:2015ipa}, G\"{o}del solution \cite{Nascimento:2014vva}, among others. In addition, studies that investigate how the quantum effects affect the bumblebee theory have been discussed. For example, the quantization and stability due to quantum effects of a vector model presenting spontaneous breaking of Lorentz symmetry have been analyzed \cite{Her} and the radiative corrections of the bumblebee electrodynamics in flat space-time have been studied \cite{Maluf:2015hda}. The main objective of this paper is to use the bumblebee gravity to determine the scale factor of the universe, considering  the dark energy density defined by the Ricci dark energy model. Then the cosmic accelerated expansion will be discussed in this context.

This paper is organized as follows. In section II, a brief introduction to the bumblebee gravity model is presented. In section III, the Friedmann-Robertson-Walker (FRW) metric is considered and the modified Friedmann equations are calculated in the  Ricci dark energy model context. The Friedmann equation is analyzed for two different cases: zero and non-zero coupling constant.  In section IV, some concluding remarks are presented.

\section{Bumblebee Model}

Here a brief introduction to the Bumblebee field is considered. The bumblebee model is a field theory that extends the standard formalism of general relativity by allowing Lorentz symmetry breaking. A vector field $B_\mu$ acquires a non-vanishing vacuum expectation value $b_\mu$ that leads to a spontaneous Lorentz symmetry breaking. The bumblebee model is among the field theories with spontaneous Lorentz and diffeomorphism violations. The spontaneous Lorentz symmetry breaking accompanied by diffeomorphism violation have been studied and it is well-know in the literature \cite{Bluhm}. The action that describes this model is 
\begin{equation}\label{action}
S=\int \sqrt{-g}\left[\frac{1}{2\kappa}(R+\xi B^\mu B^\nu R_{\mu \nu}) -\frac{1}{4}B^{\mu \nu}B_{\mu \nu}\\ -V(B^\mu B_\nu \pm b^2) +\mathcal{L}_M \right],
\end{equation}
where $g=\det(g_{\mu \nu })$ is the determinant of the metric tensor, $\kappa=8\pi G$, $R$ is the Ricci scalar, $R_{\mu\nu}$ is the Ricci tensor,  $\xi$ is the coupling constant which controls the non-minimal gravity-bumblebee interaction, $B_{\mu \nu}\equiv \partial_\mu B_\nu-\partial_\nu B_\mu$ is the field-strength tensor, $V(B^\mu B_\nu \pm b^2)$ is a potential exhibiting a non-vanishing vacuum expectation value and $\mathcal{L}_M$ is the Lagrangian density for the matter fields. The vacuum expectation value of the bumblebee field occurs when 
\bea
B_\mu B^\nu \pm b^2=0
\eea
is satisfied. This implies that the field $B_\mu$ acquires non-vanishing vacuum expectation value, i.e., $\langle B_\mu \rangle=b_\mu$ such that $ b_\mu b^\mu =\mp b^2=\rm{const.}$, then the vector background $b_\mu$ spontaneously breaks the Lorentz symmetry.  The $\pm$ signs in the potential determine whether $b_\mu$ is time-like or space-like.

The field equations are obtained varying the action, eq. (\ref{action}), with respect to the metric and to the bumblebee field. The modified Einstein equation is
\bea
G_{\mu \nu}&=&\kappa \left[2V'B_\mu B_\nu+B_{\mu \alpha}B_\nu{}^\alpha-\left(V+\frac{1}{4}B_{\alpha \beta}B^{\alpha \beta}\right)g_{\mu \nu} \right] \nonumber\\
&+&\xi \Bigl[\frac{1}{2}B^\alpha B^\beta R_{\alpha \beta}g_{\mu \nu}-B_\mu B^\alpha R_{\alpha \nu}-B_\nu B^\alpha R_{\alpha \mu}\nonumber\\ 
&+&\frac{1}{2}\nabla_\alpha \nabla_\mu(B^\alpha B_\nu)+\frac{1}{2}\nabla_\alpha \nabla_\nu(B^\alpha B_\mu) \nonumber\\ 
&-&\frac{1}{2}\nabla_\alpha \nabla_\beta(B^\alpha B^\beta)g_{\mu \nu}-\frac{1}{2}\square(B_\mu B_\nu)\Bigl]+\kappa T_{\mu \nu},
\eea
where $G_{\mu \nu}$ is the Einstein tensor, $V'$ denotes the derivative of the potential $V$ with respect to its argument and $T_{\mu \nu}$ is the energy-momentum tensor of matter. The equation of motion for the bumblebee field is given as
\begin{equation}\label{eqmov}
\nabla_\mu B^{\mu \nu}=2\left(V'B^\nu-\frac{\xi}{2\kappa}B_\mu R^{\mu \nu}\right).
\end{equation}
It is checked that when the both bumblebee field $B_\mu$ and potential $V(B_\mu)$ are vanished, the original general relativity field equations are recovered.

In the next section the RDE model will be considered in the bumblebee gravity framework.

\section{Studying the Ricci Dark Energy in Bumblebee Gravity}

We consider the Ricci dark energy in the context of bumblebee model for homogeneous and isotropic universe described by the Friedmann-Robertson-Walker (FRW) metric for flat geometries, given by
\begin{equation}\label{FRW}
ds^2=-dt^2+a^2(t)\left(dr^2+r^2d\theta ^2+r^2sin^2\theta d\phi ^2\right).
\end{equation}

The Ricci dark energy density is proportional to the scalar of curvature \cite{gaoRDE}, i.e.,
\begin{equation}\label{dRDEgeral}
\rho _X=-\frac{\alpha}{16\pi}R,
\end{equation}
where $\alpha$ is a constant to be determined. The correspondent energy-momentum tensor is
\begin{equation}\label{TRDE}
T^{RDE}_{\mu \nu}=(\rho_X+p_X)u_\mu u_\nu+p_Xg_{\mu \nu},
\end{equation}
where $p_X$ is the pressure of dark energy, given by
\begin{equation}\label{pRDE}
p_X=-\rho _X-\frac{1}{3}\frac{d\rho_x}{dx}=-\stigma a^{-\eta},
\end{equation}
where $\stigma= \left(\frac{2}{3\alpha}-\frac{1}{3}\right)f_0$, $\eta=\left(4-\frac{2}{\alpha}\right)$, $a=e^x$ and $f_0$ is an integration constant.

Considering the FRW metric, eq. (\ref{FRW}), the energy density becomes
\begin{equation}\label{dRDE}
\rho _X=\frac{3\alpha}{8\pi}\left(\frac{\ddot a}{a}+\frac{\dot a^2}{a^2}\right),
\end{equation}
where $R=-6(\dot{H}+2H^2)$ has been used. Here $H=\frac{\dot{a}}{a}$ is the Hubble parameter.
In order to calculate the field equations and maintain the validity of the cosmological principle,  the bumblebee field is chosen as
\begin{equation}\label{VBB}
B_\mu=(B(t),\vec{0}),
\end{equation}
which leads eq. (\ref{eqmov}) to
\begin{equation}\label{VlinhaB}
\left(V'-\frac{3\xi}{2\kappa}\frac{\ddot a}{a}\right)B=0,
\end{equation}
which provides the relationship between the potential dynamics and the scale factor.

The modified Friedmann equations for RDE are
\begin{align}\label{F-RDE}
H^2(1-\xi B^2)&=\frac{1}{3}\kappa(\rho_X+V)+HB\dot{B}
\\
\left(H^2+2\frac{\ddot a}{a}\right)(1-\xi B^2)&=-\kappa p_X+\xi\left[4HB\dot{B}+\dot{B}^2+B\ddot{B}\right].
\end{align}

Using the Bianchi identities, the modified equation for conservation of energy is obtained as
\begin{equation}
\dot{\rho_X}=-3H(\rho_X+p_X) -3\frac{\xi}{\kappa}\frac{\ddot{a}}{a}B(HB+\dot{B})+3\frac{\xi}{\kappa}\frac{\ddot{a}}{a}B^2.
\end{equation}
Due to the complexity of the above equations, an analytical solution for $a(t)$ and $B(t)$ can not be obtained. So we will restrict ourselves to two cases of solution. The first case where the coupling constant $\xi$ is equal to zero. And the second case we will look for a solution in the vacuum, where the bumblebee field becomes a constant that minimizes the potential.

\subsection{Zero coupling constant}

In this case $\xi=0$, then eq. (\ref{VlinhaB}) becomes
\bea
V'B=0.
\eea 
This result indicates that the bumblebee field only contributes through a constant potential $V(B^\mu B_\nu \pm b^2)=V_0$. Thus, the bumblebee field rest at one of the extremes of its potential, and keeping it from evolving with time. Then the modified Friedmann equation (\ref{F-RDE}) becomes
\begin{equation}
H^2=\frac{1}{3}\kappa(\rho_X+V).
\end{equation}
Using the energy density, eq. (\ref{dRDE}), and $V=V_0$ we get
\begin{align}
\alpha \frac{\ddot{a}}{a}+(\alpha -1)\frac{\dot{a^2}}{a^2}+c=0,
\end{align}
where $c=\frac{1}{3}\kappa V_0$. Solving this equation, the scale factor is given as
\begin{equation}
a(t)= c_2 \cos ^{\gamma}\left[\beta \left(t-\alpha  c_1\right)\right],
\end{equation}
where $\gamma=\frac{\alpha }{2 \alpha -1}$, $\beta=\frac{\sqrt{c(2 \alpha -1)}}{\alpha}$ and $c_1$ and $c_2$ are integration constants. From this result the deceleration parameter, that is defined as
\bea
q=-\frac{\ddot a}{aH^2},
\eea
is calculated and it is given as
\begin{equation}
q=\gamma^{-1} \csc ^2\left[ \beta\left(t-\alpha  c_1\right)\right]-1.
\end{equation}

To analyze the behavior of the scale factor $a(t)$ and of the deceleration parameter $q$, the $\alpha$ parameter and the potential $V_0$ should have a defined value. 

Let us consider the following case for $\alpha$ parameter and the constant potential $V_0$.

\subsubsection{$\alpha>0$ and $V_0<0$}

This case leads to a cyclic model of the universe that is composed of accelerated and decelerated phases, as shown in FIG.1. This figure displays just the real part of the functions $a(t)$ and $q$ for a given value of the parameters $\alpha$ and $V_0$. A cyclic universe has been studied in  \cite{Steinhardt,FRAMPTON201828,montani2018bianchi}.
\begin{figure}[h]
    \includegraphics[scale=0.5]{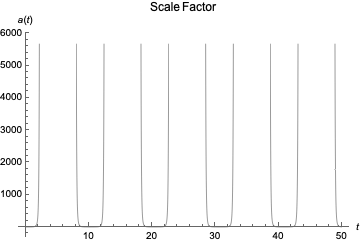}
    \includegraphics[scale=0.5]{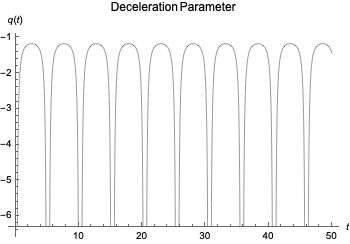}
    \caption{Evolution of the scale factor and the deceleration parameter for $\alpha>0$ and $V_0<0$.}
\end{figure}

\newpage

\subsubsection{$\alpha<0$ and $V_0>0$}

Here is recovered a framework that exhibit a universe in accelerated expansion. The FIG.2 describes this behavior.
\begin{figure}[!htb]
    \includegraphics[scale=0.5]{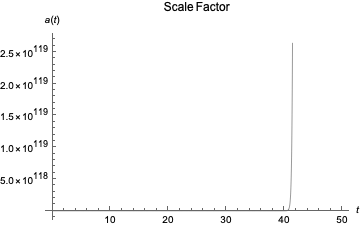}
    \includegraphics[scale=0.5]{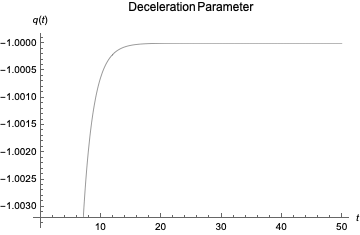}
  \caption{Evolution of the scale factor and the deceleration parameter for $\alpha<0$ and $V_0>0$.}
  \label{fig:6}
\end{figure}

An important note, in \cite{gaoRDE}, from observational data, has been determined that $\alpha=0.46$. By comparing this observational data with our result, it indicates that the universe is cyclic, i.e., it alternates between accelerated and decelerated phase during its cosmic history. However, observational data confirm that the universe is expanding in an accelerated phase. Therefore, this current acceleration of the universe may be interpreted as a phase in a cyclic universe.

\subsection{Non-zero coupling constant}

Here the vacuum solution induced by the Lorentz symmetry breaking is considered. It is possible when the bumblebee field $B_\mu$ remains frozen in its vacuum expectation value $b_\mu$, then
\bea
\langle B_\mu \rangle=b_\mu,
\eea
consequently, $V=V'=0$. Thus the particular form of the potential driving its dynamics is irrelevant \cite{Ovgun:2018xys, Casana:2017jkc, Bertolami}.

In order to obtain the vacuum solution to the bumblebee gravity let us consider a time-like background $b_\mu$ assuming the form
\begin{equation}\label{VFRW}
b_\mu=(b,\vec{0}).
\end{equation}
In this background field, all components of the corresponding field strength vanishes, i.e.,
\bea
b_{\mu\nu}=\partial_\mu b_\nu-\partial_\nu b_\mu=0.
\eea
Then the modified Friedmann equation becomes
\begin{align}\label{F-RDE1}
H^2(1-\xi B^2)=\frac{1}{3}\kappa\rho_X-\xi b^2\frac{\ddot{a}}{a}.
\end{align}
Using that $H=\frac{\dot{a}}{a}$ and eq. (\ref{dRDE}), this equation becomes
\begin{equation}\label{VFRWa}
\frac{\ddot{a}}{a}+A\frac{\dot{a}^2}{a^2}=0,
\end{equation}
where $A$ is a constant defined as
\begin{equation}\label{A}
A=\frac{\xi b^2+\alpha-1}{\alpha -\xi b^2}.
\end{equation}
The solution of eq. (\ref{VFRWa}) give us the scale factor as
\begin{equation}
a(t)=c_2 \left[(1+A)t-c_1\right]{}^{\frac{1}{A+1}},
\end{equation}
where $c_1$ and $c_2$ are integration constants. Then the deceleration parameter is given as
\begin{equation}
q=-A. 
\end{equation}

It is important to note that, if the constant $A$ is a positive number, the deceleration parameter is negative and then the universe exhibit an acceleration expansion. In addition, if the coupling constant is zero the accelerated expansion occurs just in the condition $\alpha>1$.

\section{Conclusions}

The bumblebee model is a gravitational theory that exhibit spontaneous Lorentz violation due to a vector field $B_\mu$ that acquires a nonzero vacuum expectation value. In this Lorentz-violating theory the Ricci dark energy density is considered. Then using the FRW metric the scale factor and the deceleration parameter is calculated for two different cases, zero and non-zero coupling constant. When the coupling constant is zero the bumblebee field rest at one of the extremes of its potential, and keeping it from evolving with time. In this case the scale factor and the deceleration parameter depend of a constant potential $V_0$ and of $\alpha$, a parameter that comes from the Ricci dark energy model. In addition, the choice of $\alpha$ and $V_0$ can determine a cyclic universe, with acceleration and deceleration phases, or a universe only with acceleration. In the case with non-zero coupling constant the vacuum solution is considered. This solution emerges when the bumblebee field $B_\mu$ remains frozen in its vacuum expectation value $b_\mu$. This case displays a scale factor that leads to an accelerated expansion. Therefore, our results shown that the state of accelerated expansion of the universe may be understood by a Lorentz-violating gravitational model combined with Ricci dark energy density.

\section*{Acknowledgments}

This work by A. F. S. is supported by CNPq projects 308611/2017-9 and 430194/2018-8.


\begin{thebibliography}{99}
\bibitem{Riess} A. G. Riess et al., Astrophys. J. {\bf 116}, 1009 (1998).
\bibitem{Permutter} S. Permutter et al., Astrophys. J. {\bf 517}, 565 (1999).
\bibitem{Sp1} D. N. Spergel et al., Atrophys. J. Suppl. Ser. {\bf 148}, 175 (2003).
\bibitem{Sp2} D. N. Spergel et al., Atrophys. J. Suppl. Ser. {\bf 170}, 377 (2007).
\bibitem{Teg} M. Tegmark et al., Phys. Rev. D {\bf 69}, 10350 (2004)1.
\bibitem{Se} U. Seljak et al., Phys. Rev. D {\bf 71}, 103515 (2005).
\bibitem{Will2006a} C.~M. Will, Liv. Rev. Rel. {\bf 9}, 3 (2006).
\bibitem{Spergel:2003cb} D.~N. Spergel et~al., Astrophys. J. Suppl. {\bf 148}, 175 (2003).
\bibitem{Bull:2015stt} P.~Bull et~al., Phys. Dark Univ. {\bf 12}, 56 (2016).
\bibitem{gaoRDE} C.~Gao, F.~Wu, X.~Chen, and Y.-G. Shen, Phys. Rev. D {\bf 79}, 043511 (2009).
\bibitem{Susskind} L.~Susskind, J. Math. Phys. {\bf 36}, 6377 (1995).
\bibitem{GHAFFARI201976} S.~Ghaffari,  New Astronomy {\bf 67}, 76 (2019).
\bibitem{RAO2018469} V.~Rao, U.~D. Prasanthi, and Y.~Aditya,  Results in Physics {\bf 10}, 469 (2018).
\bibitem{Salehi:2018czt} A.~Salehi, H.~Farajollahi, and S.~Aryamanesh, Grav. Cosmol {\bf 24}, 292 (2018).
\bibitem{Singh2018} C.~P. Singh and A.~Kumar, Eur. Phys. J. Plus {\bf 133}, 312 (2018).
\bibitem{Silva2013} J.~G. Silva and A.~F. Santos,  Eur. Phys. J.  C {\bf 73}, 2500 (2013).
\bibitem{PhysRevD.39.683} V.~A. Kosteleck\'y and S.~Samuel, Phys. Rev. D {\bf 39}, 683 (1989).
\bibitem{Kost1991} V. A. Kostelecky and R. Potting, Nucl. Phys. B {\bf 359}, 545 (1991); Phys. Lett. B {\bf 381}, 89 (1996).
\bibitem{Carroll} S. M. Carroll, J. A. Harvey, V. A. Kostelecky, C. D. Lane and T. Okamoto, Phys. Rev. Lett. {\bf 87}, 141601 (2001).
\bibitem{Gambini} R. Gambini, J. Pullin, Phys. Rev. D {\bf 59}, 124021 (1999).
\bibitem{SME1} D. Colladay and V. A. Kostelecky, Phys. Rev. D {\bf 55}, 6760 (1997).
\bibitem{SME2} D. Colladay and V. A. Kostelecky, Phys. Rev. D {\bf 58}, 116002 (1998).
\bibitem{Ovgun:2018xys} A. \"{O}vg\"{u}n,  K. Jusufi and I. Sakallı, {\it Exact traversable wormhole solution in bumblebee gravity},  arXiv:1804.09911 [gr-qc].
\bibitem{Casana:2017jkc} R.~Casana, A.~Cavalcante, F.~P. Poulis, and E.~B. Santos, Phys. Rev. {\bf D97}, 104001 (2018).
\bibitem{Capelo:2015ipa} D.~Capelo and J.~Paramos,  Phys. Rev. D {\bf 91}, 104007 (2015).
\bibitem{Nascimento:2014vva} A.~F. Santos, A.~{\relax Yu}. Petrov, W.~D.~R. Jesus, and J.~R. Nascimento, Mod. Phys. Lett. A {\bf 30}, 1550011 (2015).
\bibitem{Her} C. A. Hernaski, Phys. Rev. D {\bf 90}, 124036 (2014).
\bibitem{Maluf:2015hda} R.~V. Maluf, J.~E.~G. Silva, and C.~A.~S. Almeida, Phys. Lett. B {\bf 749}, 304 (2015).
\bibitem{Bluhm} R. Bluhm and V. A. Kostelecky, Phys. Rev. D {\bf 71}, 065008 (2005).
 \bibitem{capelo2015} D. Capelo and J. P\'{a}ramos, Phys. Rev. D {\bf 91}, 104007 (2015).
\bibitem{Steinhardt} P. J. Steinhardt and N. Turok, Phys. Rev. D {\bf 65}, 126003 (2002).
\bibitem{montani2018bianchi} G. Montani, A. Marchi and R. Moriconi, Phys. Lett. B {\bf 777}, 191 (2018).
\bibitem{FRAMPTON201828} P. H. Frampton, Phys. Dark Universe {\bf 20}, 28 (2018).
\bibitem{expUniv} J. A. Frieman, M. S.Turner and D. Huterer, Annual Review of Astronomy and Astrophysics {\bf 46}, 385 (2008).
\bibitem{Bertolami} O. Bertolami and J. Paramos, Phys. Rev. D {\bf 72}, 044001 (2005).
\end{thebibliography}
\end{document}